\documentclass[manuscript,screen]{acmart}
\usepackage{enumitem}
\usepackage{csquotes}
\usepackage[many]{tcolorbox}  
\usepackage{tabularx}
\tcbuselibrary{listings,breakable}
\definecolor{sub}{HTML}{ffffff}     

\tcbset{
    sharp corners,
    before skip = 0.1cm,    
    after skip = 0.01cm      
}

\def\highlightbox#1#2{
\medskip
\begin{tcolorbox}[
    enlarge bottom by=3pt,
    boxrule=0pt,
    boxsep=0pt,
    breakable,
    enhanced jigsaw,
    borderline west={4pt}{0pt}{gray},
]
#2
\end{tcolorbox}
}

\acmConference[SE 2030]{International Workshop on Software Engineering in 2030}{a}{a}

\begin{document}

\title{Towards Sustainable and Secure Reuse in Dependency Supply Chains:\\
Initial Analysis of NPM packages at the End of the Chain
}

\author{Brittany Reid}
\orcid{0000-0001-7012-0655}
\affiliation{%
  \institution{Nara Institute of Science and Technology}
  \country{Japan}
}
\email{brittany.reid@naist.ac.jp}

\author{Raula Gaikovina Kula}
\orcid{0000-0003-2324-0608}
\affiliation{%
 \institution{The University of Osaka}
 \country{Japan}}
 \email{raula-k@ist.osaka-u.ac.jp}

\renewcommand{\shortauthors}{B. Reid and R. Kula}



\setcopyright{acmcopyright}

\begin{abstract}
Much of the success of modern software development can be attributed to code reuse. The ability to reuse existing functionality via third-party dependencies has enabled massive gains in productivity, but for a long time the dominant philosophy has been to \textit{ `reuse as much as possible, without thought for what is being depended upon'}, creating fragile dependency chains. Heavy reliance has raised resiliency and maintenance concerns. In this vision paper, we investigate packages that challenge the typical concepts of reuse--that is, packages with no dependencies themselves that bear the responsibility of being at the end of the dependency supply chain. By avoiding dependencies, these packages at the end of the chain may also avoid the associated risks. Our initial analysis of the most depended upon NPM packages shows that such end-of-chain packages make up a significant portion of these critical dependency chain (over 50\%). We find that these end-of-chain packages vary in characteristics and are not just packages that can be easily replaced, and present five cases. We then ask ourselves: Should maintainers minimize external dependencies? We argue that these packages reveal important lessons for strategic reuse—balancing the undeniable benefits of dependency ecosystems with sustainable, secure practices.

\end{abstract}

\maketitle

\section{Introduction}



\noindent
Douglas McIlroy proposed the idea of mass-produced, component based software engineering in 1968~\cite{mcilroy1968mass}. He envisioned a future where software would be assembled from reusable parts rather than repeatedly built from scratch, akin to mass manufacturing. Today, over 50 years later, the mindset of `use over build' dominates the industry; almost no software is built entirely from the ground up, instead making use of vast collections of already-existing software libraries. The Node Package Manager (NPM), one of the largest ecosystems of software libraries, hosts over 3.5 million NPM `packages' as of July 2025~\cite{npm_stats},  each with their own complicated chains of dependency. Though the benefits of library use for productivity, code quality and defect reduction are well established~\cite{lim1994effects, basili1996reuse}, recent focus has shifted to the security and sustainability of software projects~\cite{decan2018impact, wattanakriengkrai2022giving}, especially in regards to resilience and maintenance. Prior work has considered code reuse as a double-edged sword~\cite{gkortzis2019double}: the maturity and open-source nature of these packages can make systems more secure, but at the same time the increased attack surface can introduce vulnerabilities. McIlroy's metaphor of industrialization, suppliers and supply chains remains just as relevant to software today as when it was proposed.

Recently, numerous high-profile cases of weaknesses in the software supply chain have gained attention from the software engineering community due to their widespread impact. The Java Log4Shell vulnerability~\cite{log4shell_surface} impacted millions of users, with 60\% of affected libraries vulnerable via indirect dependencies~\cite{snyk_log4j}. In 2016, the unpublishing of the single function \texttt{left-pad} package lead to NPM changing policies on library deletion, after large tech companies such as Meta and PayPal were left suddenly unable to build projects. Increasingly, there are also social aspects such as protestware to consider - the introduction of potentially malicious code for political or financial reasons~\cite{kula2022war}. These cases highlight the challenge posed by lengthy and complex chains of indirect dependencies that have emerged across software ecosystems such as NPM. Maintainers and users of a package may not necessarily be aware of the full scale and impact of a package's dependency chain - the result is opportunities for unsecured links on the supply chain to introduce security and maintenance issues to upstream dependent packages. Additionally, they may not be aware of who maintains the software they rely on - often open source software is maintained by small teams of volunteers. Like sourcing parts from unknown factories, the promise of mass-produced, reliable components has not been realized.

\begin{figure}[h]
    \centering
    \includegraphics[width=0.8\linewidth]{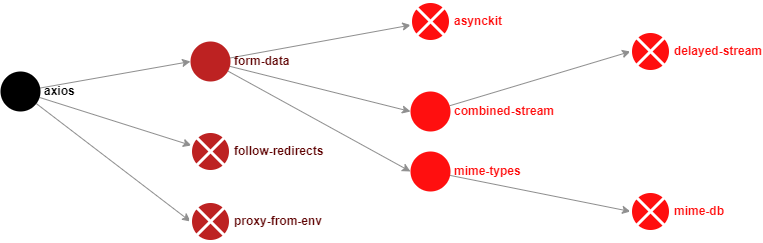}
    \caption[Example of the dependency chain for the NPM package axios.]{Example of a dependency chain for the NPM package \texttt{axios}\footnotemark. Darker nodes indicate direct dependencies, lighter nodes indicate indirect dependencies, and crosses indicate end-of-chain packages.}
    \label{fig:placeholder}
\end{figure}

\footnotetext{\url{https://www.npmjs.com/package/axios}}

Standing against these trends are projects that abstain from the use of third-party dependencies entirely. In doing so, these packages may also be avoiding the risk factors that come with dependency use. Previous work investigating the NPM ecosystem has focused on the impact of lengthy dependency chains, but not necessarily the reverse, and the results of these studies may not generalise to the no dependency case. In this vein, there may be undiscovered benefits to refraining from code reuse in specific scenarios wherein the `\textit{buck}' of the supply chain stops with these packages.

In this paper, we preform an initial analysis of the ten most depended upon libraries in the NPM ecosystem and their dependency chains (consisting of 216 unique libraries), in order to identify and characterise the types of dependencies that exist at the end of the chain. For the most popular packages on NPM, our preliminary results indicate that packages with no dependencies are indeed evident in the dependency chain, comprising just over 50\% of the dependency chain in our sample. We select a series of motivating cases, which we believe characterise some the different types of end-of-chain dependency packages that exist in the NPM ecosystem: 1) \textbf{core ecosystem packages} that are critical to the ecosystem; 2) \textbf{`classical' packages} that don't necessitate updates; 3) \textbf{unmaintained trivial packages}; 4) \textbf{`not trivial` trivial packages} that appear deceptively simple; 5) and \textbf{bundled packages} that absorb their dependencies. We believe these cases require further study and are of interest to the software engineering community.

For a long time, the dominant philosophy has been to \textit{ `reuse as much as possible, without thought for what is being depended upon'}. We propose an alternative: sustainable, secure reuse. In line with this, we pose the following overarching question: 

\highlightbox{}{
\textit{How can we strive towards more sustainable reuse, and can we learn from libraries that lack dependencies?}
}

More and more, software developers are looking to reduce dependencies and their overhead~\cite{BM25}. The Alpha and Omega Project specifically aims to help the most critical open source projects improve their security postures~\cite{AlphaOme74:online}. In this paper, we argue that analysis of software ecosystems like NPM should not ignore the role of end-of-chain packages, and how the unforeseen benefit may stack up against the productivity benefits of code reuse. We present a series of open research questions to guide researchers in investigating, characterising and understanding the perils and best practice of these no dependency packages, their impact and the motivations behind reducing dependencies, within the scope of the following topics:

\begin{itemize}[label={}]
    \item \textbf{Characterising End-Of-Chain Packages:} These questions aim to better understand the structural traits and behaviour of packages with no dependencies, including their size, update frequency, and role in the ecosystem.

    \item \textbf{Understanding Developer Motivations:} These questions explore why developers create or adopt zero-dependency packages, such as the security, performance and maintenance factors that might be behind their choices.

    \item \textbf{Evaluating the Risks, Impact and Benefits of No Dependencies:} These questions assess the reuse trade-offs of reducing dependencies on security, maintenance effort and ecosystem stability.

    \item \textbf{Rethinking Reuse for the Ecosystem:} These questions reframe reuse strategies to prioritize sustainability and resilience, with implications for governance, adoption, and long-term ecosystem health.
\end{itemize}

Overall, we envision a future where software library reuse is secure, sustainable and intentional, and where developers understand the impact and scope of their supply chains. In order to strive towards this goal, however, we must understand and study reuse from the perspective of libraries that minimise reuse. Additionally, we want to understand the role of these end-of-chain libraries on the quality and resilience of the supply chain itself, and further reveal more insights into the complicated trade-offs of code reuse in software development.

\section{How prevalent are end-of-chain libraries for the Top 10 NPM Packages?}

In order to understand the nature of end of chain libraries, we conduct an initial analysis of the top ten most depended upon NPM libraries and their dependency chains. Being one of the largest and most studied software ecosystems~\cite{npm_stats, zimmermann2019small, kula2017impact}, we choose to limit this initial analysis to only the NPM ecosystem. We select this subset of packages because they represent a sample of the packages critical to the NPM dependency chain. First, we utilise the libraries.io\footnote{\url{https://libraries.io/}} API to collect the top 10 packages sorted by dependency count. Next, for those 10 packages, we mine the NPM registry\footnote{\url{https://registry.npmjs.org}} for dependency data, and recursively build their chain of dependencies; that is, the packages they depend upon. 

For this initial analysis we look only at the \texttt{dependencies} field on the NPM registry, and ignore \texttt{devDependencies} and \texttt{peerDependencies}. \texttt{devDependencies} pertain to only dependencies required during development of the library that are not installed in production or by dependent libraries. These typically include testing frameworks and programming aids like linters. APIs from these dependencies are not called in any production code. Similarly, \texttt{peerDependencies} often denotes that a package is a plug-in for another package, but it may or may not actually use APIs from that package. We limit our analysis to regular dependencies to avoid the complexity of these cases for this investigation. Additionally, no recursive dependencies are present in this dataset. \autoref{tab:dataset} shows the final dataset statistics. 

\begin{table}[h]
    \centering
     \caption{Statistics for the total dependency chain of the top 10 most depended upon packages on NPM.}
    \begin{tabular}{lr}
    \toprule
        Root NPM packages & 10 \\
        Dependency chain size &  504 \\
        Unique packages in dependency chain & 216 \\
        End-of-Chain packages & 115 \\
        End-of-Chain packages (manually verified) & 114 \\
    \bottomrule
    \end{tabular}
    \label{tab:dataset}
\end{table}

Out of 216 packages within this dependency chain, we observe that 115 (53.24\%) have no dependencies according to the NPM registry. Additionally, five of the top ten most depended upon packages (the roots of the dependency tree) report no dependencies via this method. In \autoref{sec:cases} we take a deeper look at example cases from this sample; however, as part of our deeper analysis we also uncover that the NPM dependency data is not always consistent with the \texttt{package.json} file available on GitHub. For example, \texttt{prettier} reports zero dependencies, despite having dependencies listed in the \texttt{package.json}. However, this is the only instance in our dataset, reducing the manually verified zero dependency number to 114 packages (52.78\%). The final number of end-of-chain packages in the top ten most depended upon is four: \texttt{typescript}, \texttt{eslint-plugin-react-hooks}, \texttt{moment} and \texttt{@types/lodash}. This analysis reveals that some of the most popular packages in the NPM ecosystem have no dependencies.

\highlightbox{Observation:}{Libraries with no dependencies make up a majority of the supply chain for the most depended upon NPM packages (52.78\%), however, they remain understudied.}

\section{Motivating Examples: Exploring Five Cases of End-of-Chain Libraries}
\label{sec:cases}

From the dataset we select five motivating cases of libraries with no dependencies that exist at the end of the dependency chain, their impacts and the implications of their lack of dependencies. We select these cases from our sample based on their prospective interest to the software engineering community. 

The first case describes the case of an active, well-maintained package at the end of the dependency chain. The second case describes the case of a `classical' package that has remained unchanged for 11 years, that's lack of activity may be a side effect of its simplicity and lack of dependencies. In the third case, we take a look at the case of a trivial package nested deep in the dependency chain of the popular \texttt{eslint} package. Forth, we observe the case of a package that may appear trivial, but package history and activity suggests otherwise. Finally, we look at the case of a package that bundled up and absorbed its dependencies.


\subsection{Case 1: The Core Ecosystem Package}
\label{sec:case1}

The most depended upon NPM package according to \texttt{libraries.io} data also has zero dependencies: the \texttt{typescript} package\footnote{\url{https://github.com/microsoft/TypeScript}}. Given the popularity of the TypeScript language, this isn't surprising. The package is often installed as part of the build process for a package, where TypeScript will be transpiled into JavaScript. As of July 2025, 1,063,224 of the 3,583,585 libraries on the NPM registry make use of the \texttt{typescript} library - almost 1/3 of the entire ecosystem. This makes \texttt{typescript} a critical part of infrastructure in the NPM ecosystem. 

The GitHub repository for the \texttt{typescript} package is extremely active, with recent commits, 5000+ issues, 338 pull requests and a large list of 761 contributors. As the package is maintained by Microsoft, there is also an expected level of support. In terms of security, the repository has a security policy and no reported vulnerabilities on Snyk~\cite{snyk_ts}. A look at the history of this package's dependencies reveals that it has always had zero dependencies. Comparatively, the library's unpacked size is relatively large at 32MB (when most packages are measured in KB).

\highlightbox{Observation:}{
Some libraries that play an important role in the NPM ecosystem, such as \texttt{typescript}, have no dependencies. Due to their position in the Node.js ecosystem, such core ecosystem libraries may be more likely to have smaller supply chain footprints as a conscious decision to reduce supply chain risk and maintain more control. Additionally, it may be that libraries that remain self-contained are more likely to see widespread adoption by software developers. Additionally, it's size and complexity show that even large projects can avoid dependencies. These observations raise important questions about the \textbf{motivations} and \textbf{characteristics} of end-of-chain libraries. \textbf{Core Ecosystem Packages are likely to be characterised by active maintenance, high dependent count and large communities.}}

\subsection{Case 2: The `Classical' Package}
\label{sec:case2}

The \texttt{imurmurhash}\footnote{\url{https://github.com/jensyt/imurmurhash-js}} package is an interesting case; an implementation of the MurmurHash3 hashing algorithm, last updated 11 years ago, with a single maintainer. Despite these factors, it still has 23.6 million users and an active maintainer. Due to the nature of the library - an implementation of an algorithm to a specification - there is little need for the kinds of updates other libraries require, especially with no dependencies to update. In some sense, these libraries may be 'classical'.

There is only a single, closed issue concerning a false positive virus detection~\cite{imurmurhash_issue}.

\begin{displayquote}
\textit{Well, guys... Nothing has happened here in three years and the repo here hasn't been updated in 10 (!) years.
I would say it's time to stop using this package and let everyone who uses it know about it.}
\end{displayquote}

The maintainer's response, as of 2023, is simple:

\begin{displayquote}
\textit{The repo hasn't been updated because there's nothing to update, unfortunately.}
\end{displayquote}

\highlightbox{Observation:}{
Some libraries may decrease the need for maintenance by avoiding dependencies. Additionally, there may be cases of libraries that are straightforward to implement without dependencies. This case raises interesting questions concerning the \textbf{risks and benefits} of using no dependencies, as well as insights about what kinds of libraries can avoid dependencies. \textbf{Classical Packages are likely to be characterised by lack of recent updates, but maintainers that respond to issues and no active vulnerability reports for the latest version.}
}

\subsection{Case 3: The Unmaintained Trivial Package}
\label{sec:case3}

Trivial packages are small NPM packages that implement a single task. Recent research has investigated the consequences of relying on these packages along the dependency chain~\cite{abdalkareem2017why}. We look at the case of \texttt{is-number}\footnote{\url{https://github.com/jonschlinkert/is-number}}, one such trivial package with a simple task: checking if a value is a number. The package has 25.7 million users on GitHub, however it only has four contributors and hasn't been updated in 6 years. This tiny package is part of the dependency chain of the popular \texttt{eslint} package. We present this case in contrast to the classical package case above.

With a number of open pull requests, an issue was opened urging the maintainer to merge some changes relating to optimisation, especially due to the popularity of the package~\cite{isnumber_issue}.

\begin{displayquote}
    \textit{This package has 69 million weekly downloads. Please merge the optimisation pull requests.}
\end{displayquote}

When one commenter asks why anyone even uses the package for such a simple task, another comments:

\begin{displayquote}
    \textit{{[The package]} was in a package-json of mine. eslint-config-next uses is-number in a deeply nested dependency. ESLint might run 0.1\% faster if the optimisation was implemented.}
\end{displayquote}

Whether the optimisation changes would be impactful or not, this package is clearly a case of the dependency chain relying on a trivial, inactive package with only a few contributors. Unlike the previous case of a classical package, however, there are possibly needed updates. 

\highlightbox{Observation:}{
    Not every library at the end of the chain is useful; in fact, unmaintained trivial packages like \textit{is-number} may be an obvious candidate for removal in projects looking to reduce dependencies themselves. Libraries such as \textit{is-number} also raise questions about the types of libraries that can avoid dependency use. \textbf{Unmaintained trivial packages are likely to be characterised as trivial packages with no recent commits, inactive maintainers and stale issues and/or pull requests.}
}

\subsection{Case 4: The 'Not Trivial' Trivial Package}
\label{sec:case4}

Some packages may appear trivial in that they solve a single problem, have a small project size or simple, one function API, typically meeting the requirements for a `trivial' or `micro' package, however, the task they deal with is deceptively complex. One common target area for software attacks is user input, and thus, functionality that manipulates strings or relies on regular expressions are often vectors to introduce vulnerabilities. packages that deal in these areas may be intrinsically more vulnerable to outside attack, necessitating more frequent updates compared to other cases of simple packages. The \texttt{ms}\footnote{\url{https://github.com/vercel/ms}} package is a package to convert string times to milliseconds. Like other trivial packages, this package is small (6.72kB) and performs a single task. However, a deeper look into the package reveals a different story: there are 36 contributors, 17 open issues and 12 pull requests, with the last commit 7 months ago. Additionally, Snyk reports two security vulnerabilities in past versions of the package: regular expression denial of service across two ranges of versions~\cite{snyk_ms}. Despite this, the package has 31.1 million users on GitHub.

\highlightbox{Observation:}{
This package is an example of how small, seemingly simple packages can introduce weak points in the dependency chain. Due to the single functionality, lack of dependencies and small size, this package may appear deceptively trivial to potential users. The use of these types of packages may not necessarily be unwarranted though; by reusing a package, vulnerabilities can be reported, consumers of a package can be notified, and updates can be pushed. This package shows the \textbf{risks and benefits} of code reuse, in regards to vulnerabilities, but also community reporting of those vulnerabilities. \textbf{`Not trivial' trivial packages are likely to be characterised as trivial packages with active maintenance, active community (issues and PRs), history of vulnerabilities and high contributor counts.}
}

\subsection{Case 5: The Bundled Package}
\label{sec:case5}

We present another interesting case, where a zero dependency package absorbed its dependencies. The \texttt{chalk} package\footnote{\url{https://github.com/chalk/chalk}}, for styling terminal text using ANSI, has no dependencies, but this was not always the case. Prior to removing all dependencies, the package had two dependencies: \texttt{supports-color} and \texttt{ansi-styles}, both also maintained under the GitHub chalk profile. In 2021, the packages were bundled; their files were moved into the `vendor' folder of the \texttt{chalk} package~\cite{chalk_commit}, while the dependencies are still maintained separately. In the GitHub issue discussing the roadmap for this update, the main contributor explains their reasoning~\cite{chalk_issue}:

\begin{displayquote}
     \textit{The benefit is require speed and less dependencies, so faster installs. Usually, this would not be a problem, but Chalk is just used so much. Chalk is for Node.js. The consumer would never bundle themselves.} 
\end{displayquote}

Multiple commenters expressed their concerns about bundling dependencies and its impact on the NPM ecosystem, for example:

\begin{displayquote}
\textit{{[Bundling]} has a number of downsides: No vulnerability reports upon install, no deprecation reports upon install (and no letting npm automatically select versions for you in the event something gets deprecated), guaranteed n * m install size were n is the number of different versions of Chalk within a large project (could be many, many times) and m is the number of dependencies Chalk has ... I'm sure there are other downsides that I'm missing but it all feels like a hack and a move backwards. I'm not convinced either.}
\end{displayquote}

One user even calls into question if the dependencies should be separate to begin with:

\begin{displayquote}
\textit{If they are really not used by anyone else, why are they separate packages in the first place? I would find it concerning if packages start adopting this approach of bundling deps. It's npm's job to resolve dependencies, and only the top-level consumer should bundle. Some other concerns have been mentioned like what if there is a security vulnerability reported, and npm audit won't report it, and nobody has the ability to upgrade through a soft range. What if the top-level consumer bundles, and only uses a fraction of chalk API (and transitive dependencies), but tree shaking cannot kick in because chalk and its deps are already bundled.}
\end{displayquote}

While the developer decided not to go through with bundling at the time, they later did so anyway, though further discussion cannot be found on the repository. 

\highlightbox{Observation:}{
Cases like these may help us understand how and when developers choose to reduce dependencies, and specifically, when they choose to bundle over other possibilities. The frequency of this technique is currently unknown, as are its risks and benefits. \textbf{Bundled packages are likely to be characterised as packages with a previous history of dependencies, with dedicated directories such as `vendor'.}
}

\section{Discussion}

Our results suggest that end-of-chain packages potentially play an important role in the NPM ecosystem, with four of the ten most depended upon packages on the dependency chain having no dependencies themselves. As summarized in Table \ref{tab:taxonomy}, we identify five categories of end-of-chain libraries, their importance to the software ecosystem and researchers, and potential indicators. While there are obvious benefits to a lack of dependence on other software, such as a lack of inherited vulnerabilities and smaller project footprint, other benefits remain unknown. Additionally, we don't know how these gains stack up against the productivity benefits of code reuse. By looking at five example cases of packages with no dependencies at the end of the supply chain, we make the case that there may be new insights to be gained from similar packages across software ecosystems. Additionally, End-of-chain packages may have their own risk that need to considered, and the motivations behind the use of no dependencies are currently unstudied. We present a research agenda for understanding these aspects of end-of-chain packages.

\begin{table}[t]
    \centering
    \caption{Potential taxonomy of the end-of-chain libraries }
    \begin{tabular}{|l|p{50mm}|p{50mm}|}
        \hline
        \textbf{Category} & \textbf{Importance} & \textbf{Possible Characteristics}\\
        \hline
        Core Ecosystem Packages & Packages that are critical to the ecosystem. & Active Maintenance\newline High Dependent Count \newline Large Community \\
        \hline
        Classical Packages &  Packages that remain stable, but may be misidentified as inactive. & No Recent Commits\newline 
        Responsive Maintainer\newline No Active Vulnerability Reports\\
        \hline
        Unmaintained Trivial Packages & Unmaintained trivial packages that might be candidates for removal. & Small Size\newline No Recent Commits\newline No Maintainer Activity\newline Open Issues and PRs\\
        \hline
        Not Trivial Trivial Packages & Packages that solve deceptively simple tasks and require care. &  Small Size\newline Active Maintenance\newline Issue Activity\newline History of Vulnerabilities\newline High Contributor Count\\
        \hline
        Bundled Packages & Packages that chose to remove dependencies and instead bundle. & History of Dependencies\newline Existence of `vendor' Directory.\\
        \hline
    \end{tabular}
    \label{tab:taxonomy}
\end{table}

Answering the proposed research questions will take multiple different approaches. Firstly, the systematic analysis of a large dataset of packages across multiple ecosystems, including NPM, but also PyPI and Maven.As our current observations may not be generalisable to other software ecosystems, we hope to look beyond just NPM packages: different ecosystems have different policies and communities that could reveal different insights. To understand the motivations behind end-of-chain packages, we need to seek the perspectives of the developers behind these packages, via developer surveys, interviews or analysis of GitHub discussions. Ultimately, our hope is that by answering these questions, we can better understand when understanding of the concept of reuse.

\section{Research Agenda}
\label{sec:agenda}

Our initial analysis reveals five cases of packages that do not follow the majority mindset of `reuse as much as possible', but bring forth interesting questions about each library's position in the software ecosystem. Informed from these cases, we split the research agenda into four dimensions to further investigate the case of end-of-chain packages.

\subsection{Characterising End-of-Chain Packages}
The characterization of end-of-chain packages is crucial to understanding their role in the software ecosystem and the strategic decisions around their reuse. These packages, with minimal dependencies, provide unique insights into how reuse can be optimized for stability and resilience in software projects. The aim is to explore the behavioural and structural characteristics of end-of-chain packages, such as size, update frequency and popularity. 

\begin{enumerate}
    \item \textbf{Are core, highly-used packages more likely to have no dependencies?} 
    As we discovered with \hyperref[sec:case1]{Case 1}, some of the most popular packages in the NPM ecosystem have no dependencies. We aim to assess whether there is a statistical correlation between popularity measures such as download count, dependent count, etc. and having no dependencies.
    
    \item \textbf{What types of functionality are often implemented by end-of-chain packages?} 
    Many of the cases we investigated involved trivial, single purpose packages. These simple, limited scope libraries may lend themselves to having less dependencies. We ask whether certain problem domains are particularly likely to result in dependency-free implementations, either due to simplicity or to reduce attack surface.
    
    \item \textbf{Do end-of-chain libraries always begin with no dependencies, or do they emerge from reduction efforts?} 
    Some packages may have once used dependencies that were later removed. We seek to track their history to understand how and why dependency reduction occurs over time.
    
    \item \textbf{What techniques do developers use to reduce dependencies, and how prevalent are they (e.g. bundling, reimplementation)?} 
    We aim to understand the common techniques for reducing dependencies, based on the histories of existing end-of-chain packages.
\end{enumerate}

\subsection{Understanding Developer Motivations}
In this section, we aim to delve deeper into the motivations of developers that both maintain or adopt end-of-chain libraries. The decision to eliminate dependencies may be influenced by concerns over security, package size, performance, or long-term maintainability. By understanding these choices, we can better frame the cultural, technical, and risk-based reasoning behind reuse decisions and ultimately provide more informed guidance for sustainable package development and adoption.

\begin{enumerate}[resume]
    \item \textbf{What factors motivate developers to avoid dependencies?}
    Developers may avoid dependencies to reduce security risks, minimise bloat, simplify maintenance, or retain control over code. We aim to investigate which motivations are most common and how they vary across contexts.
    
    \item \textbf{When and why do developers choose to bundle dependencies?}
    In \hyperref[sec:case5]{Case 5} we observe an example of bundling, with cited motivations such as speed. However, it may also obscure origin and trust boundaries. We aim to investigate when and why developers make these trade-offs.
    
    \item \textbf{To what extent do developers consider a package's dependency chain when selecting what package to adopt?}
    Are adoption decisions influenced by the presence or absence of dependencies, and if so, under what conditions? We explore whether dependency count plays a measurable role in decision-making for package adopters.
\end{enumerate}

\subsection{Evaluating the Risks, Impact and Benefits of No Dependencies}
Removing dependencies can have risks and benefits. This section investigates the ecosystem-level implications of depending on—or eliminating—end-of-chain packages.

\begin{enumerate}[resume]
     \item \textbf{What impact does removing or replacing end-of-chain packages have on the broader software ecosystem, and how should reuse strategies account for this?}
     We aim to investigate how the removal or replacement of end-of-chain packages affects projects that depend on them. Looking at the example in \hyperref[sec:case3]{Case 3}, we can observe that some end-of-chain libraires may be candidates for removal. We can study how upstream packages can be affected and whether the functionality can be easily absorbed.

    \item \textbf{Do end-of-chain packages tend to require less long-term maintenance?} 
    Without dependency churn, these packages may require fewer updates (as in \hyperref[sec:case2]{Case 2}). We examine whether this correlates with reduced maintenance cost over time.
    
    \item \textbf{Are end-of-chain packages more secure?} 
    Having no dependencies may reduce the attack surface, but this alone may not guarantee safety. We investigate whether these packages demonstrate lower vulnerability rates in practice.
    
    \item \textbf{What are the risks and benefits of bundling dependencies, and how can this inform reuse strategy?}
    Bundling may improve resilience and performance, but also reduce transparency. We examine its tradeoffs in real-world ecosystems. We aim to develop decision criteria that developers can use to decide whether to absorb or remove dependencies, taking into account long-term reuse and minimal dependency management.
    
    \item \textbf{When should developers reuse a package versus reduce dependency?}
    Based on our taxonomy, we aim to identify cases where reuse adds clear value versus when it adds unnecessary risk or overhead, in order to optimize the decision makjing process for developers.
\end{enumerate}

\subsection{Rethinking Reuse for the Ecosystem}
To rethink reuse, it is essential to examine the role of end-of-chain libraries, when and why dependencies should be reduced, and the risks and benefits. Informed by the previous research questions, we propose a redefining of reuse strategies that that prioritize resilience and minimize risk, in order to enhance the overall health of the ecosystem. This redefinition can lead to more efficient, secure, and sustainable software architectures in the long run. The following questions focus on measuring impact, governance, and creating feedback mechanisms that foster sustainable reuse.

\begin{enumerate}[resume]
    \item \textbf{Can we develop a new model for package reuse that prioritizes self-sufficiency and reduces upstream dependency risk?}
    We aim to propose a new model for package reuse that encourages self-sufficiency, reduces risk exposure from upstream vulnerabilities, and focuses on reducing unnecessary dependencies. We will explore how end-of-chain packages can serve as examples for creating more autonomous software components in the ecosystem.

    \item \textbf{What factors contribute to the long-term sustainability of packages at the end of the chain, and how can this inform reuse policies?}
    Study the factors that enable the long-term sustainability of end-of-chain packages, such as their lack of dependencies, community support, and focus on stability. Develop best practices for maintaining and reusing such packages over extended periods, considering both their technical and community aspects.

    \item \textbf{How can we measure the impact of reduced dependency on both the quality and performance of reused packages?}
    Develop performance and quality metrics for evaluating packages that reduce their dependencies. Study how reducing dependencies can improve performance, security, and maintainability, and create guidelines for balancing these metrics when deciding to reuse a package.

    \item \textbf{What are the governance implications of shifting to a minimal dependency, end-of-chain package model in terms of community support and project lifecycles?}
    Approach: Investigate the governance implications of prioritizing minimal dependency models. How does this affect project lifecycles, developer participation, and long-term community support for these packages? Propose governance structures that encourage the adoption of minimal dependency practices.
\end{enumerate}

\section{Conclusion}

In this paper, we identify the prevalence of end-of-chain packages within the dependency chain of the most commonly used NPM libraries, making up over half of dependencies in our sample. Through five motivating cases, we show that these packages are not homogeneous: some are core ecosystem infrastructure with active maintenance, some remain stable due to their simplicity, while others raise questions about triviality, risk, and sustainability. Together, these examples illustrate that dependency avoidance is not merely an accident of design, but can also be a conscious strategy.

We argue that understanding end-of-chain packages is essential to rethinking software reuse. Whereas the prevailing philosophy of “reuse as much as possible” has accelerated development, it has also introduced fragility into software supply chains. End-of-chain packages challenge this narrative by suggesting that in certain contexts, minimizing or eliminating dependencies can lead to greater sustainability and control. At the same time, these packages present their own trade-offs and potential risks, which remain understudied.

Looking forward, we present a clear research agenda: characterizing the traits of end-of-chain libraries across ecosystems, uncovering developer motivations for reducing dependencies, and evaluating the broader trade-offs for security, ecosystem health, and sustainable reuse. By doing so, we can move toward practices of reuse that are more intentional, balanced, and resilient—where developers not only benefit from shared code, but also understand the limits and responsibilities that come with dependency choices.

\bibliographystyle{ACM-Reference-Format}
\bibliography{references}
\end{document}